\documentstyle[twocolumn,aps,psfig,floats]{revtex}  

\begin{document} 

\def\be{\begin{equation}}
\def\ee{\end{equation}}

\draft  

\twocolumn[\hsize\textwidth\columnwidth\hsize\csname %
@twocolumnfalse\endcsname

\title{Multiscale Random-Walk Algorithm for 
Simulating Interfacial Pattern Formation} 

\author{Mathis Plapp and Alain Karma}

\address{
Physics Department and Center for Interdisciplinary Research
on Complex Systems, \\
Northeastern University, Boston, Massachusetts 02115 
}

\date{June 23, 1999}

\maketitle

\begin{abstract}
We present a novel computational method
to simulate accurately a wide range of interfacial patterns
whose growth is limited by a large 
scale diffusion field. To illustrate the computational power
of this method, we demonstrate
that it can be used to simulate three-dimensional dendritic growth
in a previously unreachable range of low undercoolings 
that is of direct experimental relevance.
\end{abstract}

]

Interfacial patterns
form spontaneously in a wide range of physical
and biological systems where the motion of an
interface is limited by one (or several) diffusion fields,
each one obeying the diffusion equation
\begin{equation}
\partial_tu=D\nabla^2 u\label{e1}
\end{equation}
with specified boundary conditions on the interface.
Classic examples include various 
cellular, dendritic, or eutectic solidification
patterns (where $u$ represents 
the temperature or an impurity concentration) \cite{Kurz},
dendrite-like branched patterns 
formed during electrochemical deposition (where $u$ 
is some ion concentration) \cite{eldep}
and complex growth morphologies produced by 
bacterial colonies under stress (where $u$ can represent
the concentration of some nutrient or a
signaling agent) \cite{BenJacob}.

When attempting to accurately simulate the
growth of such structures, one is
generally faced with two major difficulties.
The first one is front tracking, which requires to
resolve accurately the boundary conditions imposed on $u$ 
at the evolving interface.
Dendritic solidification, where even a 
weak crystalline anisotropy crucially
influences the morphological development epitomizes this
difficulty. The second problem is the large disparity of scale
between the growing structure and the diffusion
field surrounding it. The physical origin
of this disparity is essentially dimensional.
The diffusion field decays ahead of the
growth structure on a length scale 
$l\sim D/v$, where $v$ is the velocity of the advancing 
interface, whereas the characteristic scale of 
the structure, e.g. the tip radius $\rho$ of a growing dendrite,
is itself the geometric mean of $l$ and a short length scale
cutoff proportional to the interface thickness. 
Thus, for small growth rate, $l$ can be
several orders of magnitude larger than $\rho$,
and simulations that resolve simultaneously the
details of the interfacial pattern and the 
diffusion field become extremely difficult.

Whereas various methods have been successfully
developed to handle front tracking,
bridging the length scale gap between
$\rho$ and $l$ has remained a
major computational challenge.
A natural idea to overcome this problem
is to use multigrid or adaptive mesh 
refinement algorithms that make the grid progressively
coarser away from the interface \cite{Sch,BraMur,Proetal98}.
However, such methods need to dynamically adapt their
grids to follow the moving interface. This is a
nontrivial task, and quantitative simulations 
of dendritic crystal growth at low undercooling have remained restricted to 
two dimensions (2-d) \cite{Proetal98}.

In this letter, we present a novel hybrid
computational approach
that efficiently bridges this length scale gap.
Over most of the computational domain, the diffusion
equation is simulated by an ensemble 
of off-lattice random walkers that take 
longer, and concomitantly rarer, steps with increasing
distance away from the growing interface.
This drastically reduces the computational cost for
evolving the large-scale field.
Moreover, a short distance away from the interface, 
this stochastic evolution is connected to
a finite-difference deterministic solution of the
interface evolution. This conversion, in turn, reduces 
the inherent noise of the stochastic method
to a negligibly small level at the interface.
This approach is relatively simple to implement in
both 2-d and 3-d while being at the same time quantitatively accurate.
Here we sketch the method and then report results
that demonstrate its capability to yield new
quantitative predictions testable by experiments in
the context of dendritic crystal growth. 
For clarity, we expose the method 
in this context although it will become clear below that
it is general. 

Let us consider a solid-liquid interface whose 
motion is limited by heat diffusion and define a
scaled temperature field $u$ that is zero in equilibrium
and equal to $-\Delta$ in the 
liquid far from the interface, where $\Delta$ is the
dimensionless undercooling. 
At the interface, $u$ satisfies the well-known boundary conditions
\begin{eqnarray}
v_n&=&D\left(\left.\partial_n u\right|_s
-\left.\partial_n u\right|_l\right),\label{e2}\\
u&=&-d_0\sum_{i=1}^2\left[a({\hat n})
+\partial_{\theta_i}^2a({\hat n})\right]\kappa_i,\label{e3}
\end{eqnarray}
corresponding to heat conservation and local 
thermodynamic equilibrium at the interface,
respectively, where $v_n$ is the 
normal velocity of the interface, $d_0$ is a microscopic
capillary length, $\theta_i$ are the local angles between
the normal $\hat n$ to the interface and the two local principal
directions on the interface, $\kappa_i$ are the 
principal curvatures, and the function $a(\hat n)$
describes the orientation dependence of 
the surface energy.

\begin{figure}
\centerline{
\psfig{file=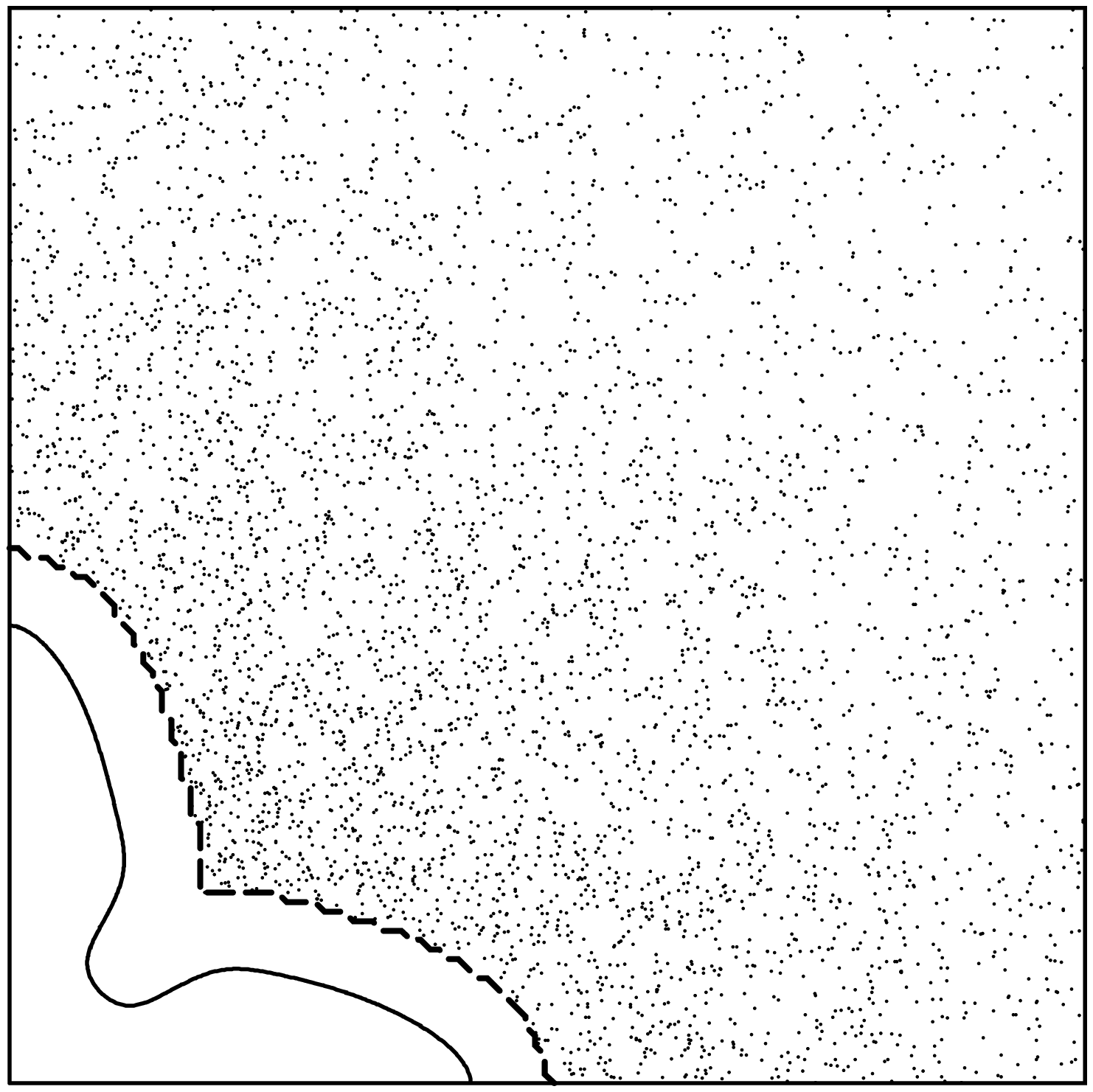,width=.3\textwidth}}
\centerline{
\psfig{file=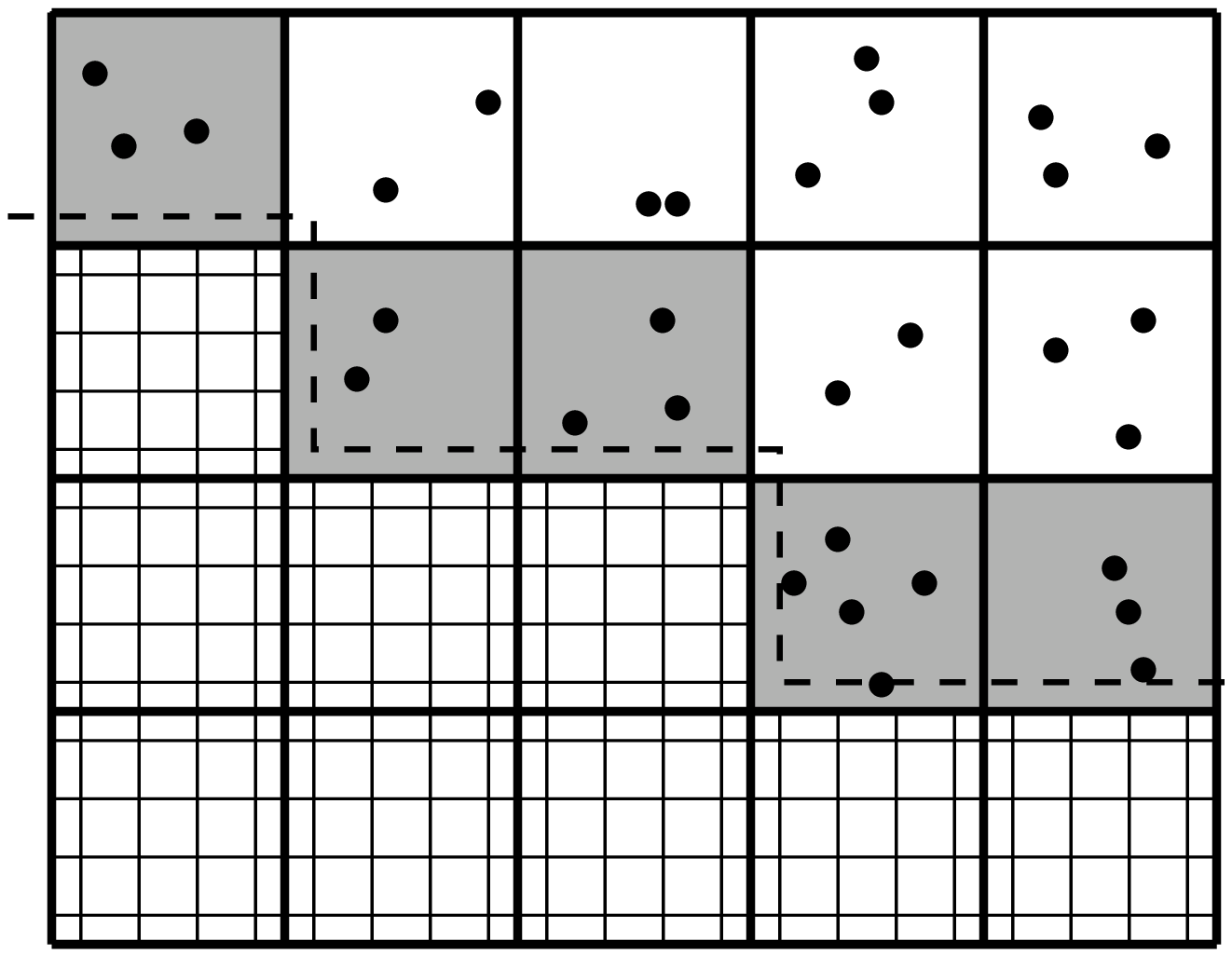,width=.275\textwidth}}
\smallskip
\caption{Top: snapshot of a 2-d simulation
for $\Delta=0.1$ and $\epsilon_4=0.025$ showing
the solid-liquid interface (solid line), 
the inner-outer domain conversion boundary (dashed-line), 
and the random walkers (dots). Only a small part of the
outer domain and one out of 50 walkers
are shown for clarity. Bottom: enlarged view
of the conversion boundary (dashed line) showing the fine
and coarse grids. Shaded cells are conversion
cells and walkers (dots) are restricted to the
outer region. 
}
\label{figden1}
\end{figure}

The basic idea of the method is to divide
space into an `inner'
and an `outer' domain as illustrated
in Fig. \ref{figden1}.
The inner domain consists of the 
growing structure and a thin `buffer layer' of liquid
surrounding the interface. The outer
domain corresponds to the rest of the liquid, and
is much larger than the inner domain at low undercooling.
In the inner domain, we solve 
{\it deterministically} the diffusion equation
on a fine uniform mesh. Moreover, for the present
crystal growth application, we handle front
tracking using a phase-field approach \cite{pfintro}, and time-step
explicitly both $u$ and the phase-field in the
inner region using the same procedure as Karma and Rappel \cite{KarRap}.
The geometry of Fig. \ref{figden1}, however, 
implies that any other front tracking method 
that solves the diffusion equation
on a uniform mesh could be used instead. Moreover,
since the boundary conditions on $u$ need not necessarily be those
defined by Eqs. \ref{e2} and \ref{e3}, the method 
obviously extends to other diffusion-limited 
pattern forming systems.

In the outer domain, the diffusion equation 
is simulated {\it stochastically} by an ensemble of
off-lattice random walkers.
The idea of solving the diffusion or Laplace equation
with an ensemble of random walkers is well-known and has
been used previously to simulate diffusion-limited growth
\cite{Vic} and Hele-Shaw flow \cite{Kad}.
The main new feature of our method
is that we have separated the 
solid-liquid interface from the boundary at which the
conversion from the deterministic to the stochastic solution
of the diffusion equation takes place.
This separation yields two essential 
benefits. Firstly, it makes it possible
to use the phase-field approach with its proven accuracy 
to simulate the interface evolution and to resolve even a weak crystalline
anisotropy, without being affected
by the details of the conversion process. 
Secondly, in the buffer layer between the solid-liquid interface and
the conversion boundary, the temperature obeys the deterministic
diffusion equation. Consequently, the noise created by the 
stochastic release and impingement of walkers is rapidly damped
away from the conversion boundary.
Hence, the amplitude of temperature fluctuations
{\em at the solid-liquid interface} can be reduced to an
insignificant level without much cost in computation
time by increasing the thickness of the buffer layer.

To connect the inner (deterministic) and outer (stochastic)
solutions, we have to supply a boundary condition
for the integration of the inner region, and we must
specify how walkers are created and absorbed at the
boundary between inner and outer regions.
Both processes are handled by using a coarse-grained
grid that is superimposed on the fine grid of the
inner region as shown in
Fig. \ref{figden1}. Cells of the coarse grid
on the border between inner and outer region,
shaded in Fig. \ref{figden1}, are called
conversion cells. The temperature in a conversion 
cell is related to the local density of walkers,
\be
u_{cc} = -\Delta\left(1-{m_i(t)/M}\right),
\ee
where $m_i(t)$ is the number of walkers in conversion
cell number $i$ at time $t$, and $M\gg 1$ is a fixed integer.
Hence, an empty cell corresponds to $u=-\Delta$ (the initial
state), whereas a box containing $M$ walkers corresponds to $u=0$. 
This formula is used in each timestep to obtain the boundary 
condition for the integration of the inner region. 
Next, we determine the quantity of heat that flows
in or out of each conversion cell from the inner
region, and add this amount to a reservoir variable
$H_i(t)$ which describes the heat content of conversion 
cell number $i$. If this variable exceeds a critical
value $H_c$, a walker is created and $H_c$ is
subtracted from the reservoir. Conversely, if $H_i(t)$
falls below $-H_c$, a walker is absorbed and $H_c$ is
added to the reservoir. This procedure assures that
walkers are created and absorbed at a rate which is
proportional to the local heat flux, and each walker
corresponds to the same discrete amount of heat.

Evidently, as the structure grows the geometry of the
conversion boundary and the configuration of the
coarse grid need to be periodically updated in order to
maintain a constant thickness of the liquid buffer
layer. This procedure, however, is straightforward
since the structure of the grids does not change. 

In the outer region, each walker is represented by a set of
variables indicating its position and the time it has next to
be updated. To update a walker, a new position is randomly
selected with a probability distribution given by the
diffusion kernel,
\begin{equation}
P(\vec x' ,t'|\vec x,t) = 
   \frac{1}{\left[4\pi D(t'-t)\right]^{d/2}}
     \exp -{{|\vec x' -\vec x|}^2\over 4D(t'-t)},
\end{equation}
where $\vec x$ and $\vec x'$ are the old and
new positions of the walker, respectively, and $t'-t$ is
the time increment between updates. This representation
of the diffusion equation is widely 
used in quantum Monte Carlo methods \cite{KooMer}. 
The key improvement that makes the algorithm efficient 
in the present context is the introduction of a variable
step size: we allow walkers to take progressively larger 
steps with increasing distance away from the interface,
and to be concomitantly updated more rarely, which does not
affect the quality of the solution near the solid-liquid 
interface. Adaptive steps have been
previously used to speed up simulations of
diffusion-limited aggregation, albeit in a simpler
Laplacian context where the stepping time is irrelevant
and only one walker at a time is simulated \cite{Meakin}.
Typically, we vary the average step size
between a value comparable to the spacing of the
inner mesh to about $100$ times that value.
Our test dendrite computations show that the far
field can be evolved at essentially no 
extra cost: the program spends most of its time
in the inner region and for the walkers which are near to
the conversion boundary and have to make small steps.
Thus this `adaptive step' 
implementation yields essentially 
the same benefits as an adaptive meshing
algorithm, while avoiding the overhead 
of regridding. Finally, our method can be
easily parallelized as will be discussed in
more details elsewhere.

\begin{figure}
\centerline{
\psfig{file=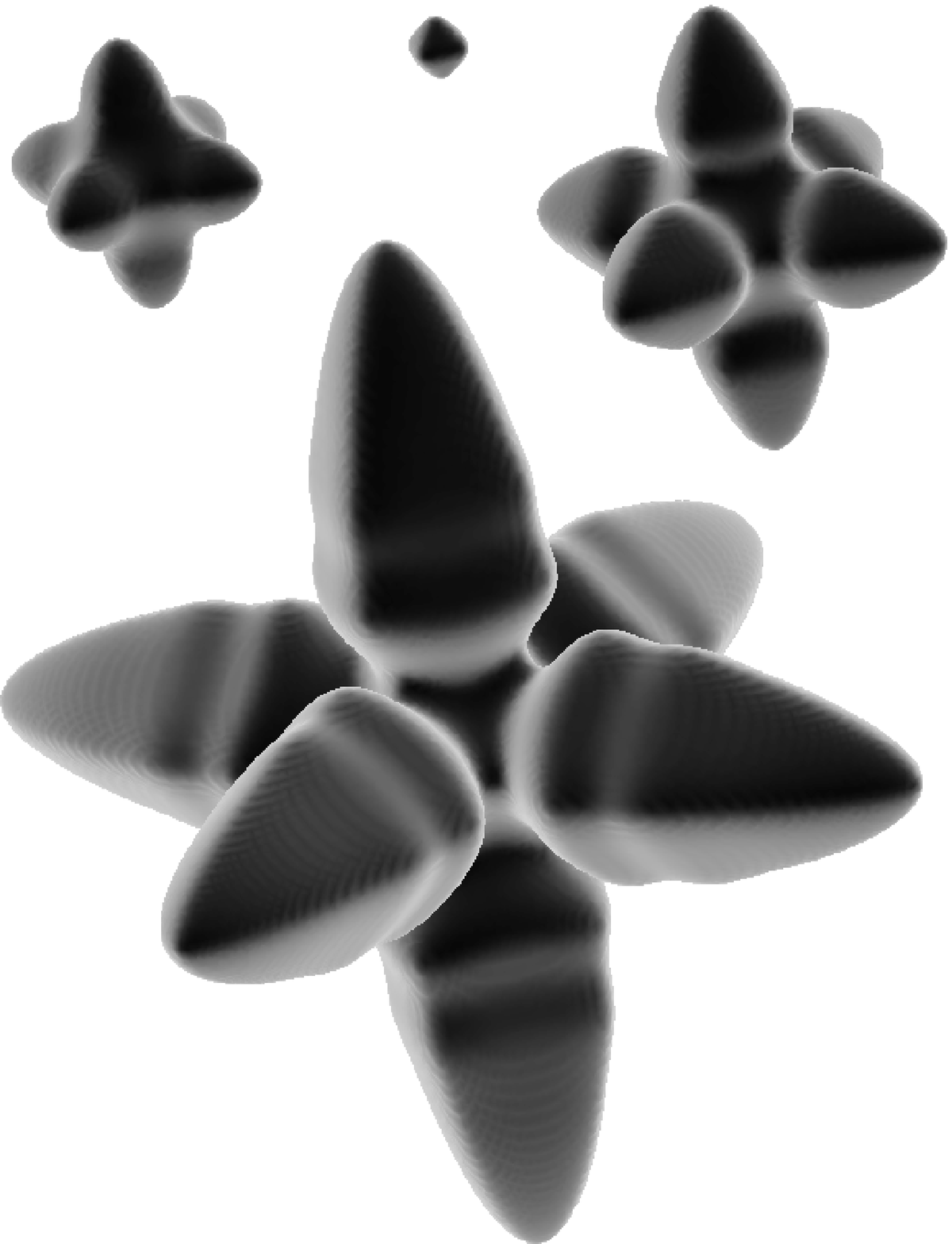,width=.175\textwidth}}
\smallskip
\centerline{
\psfig{file=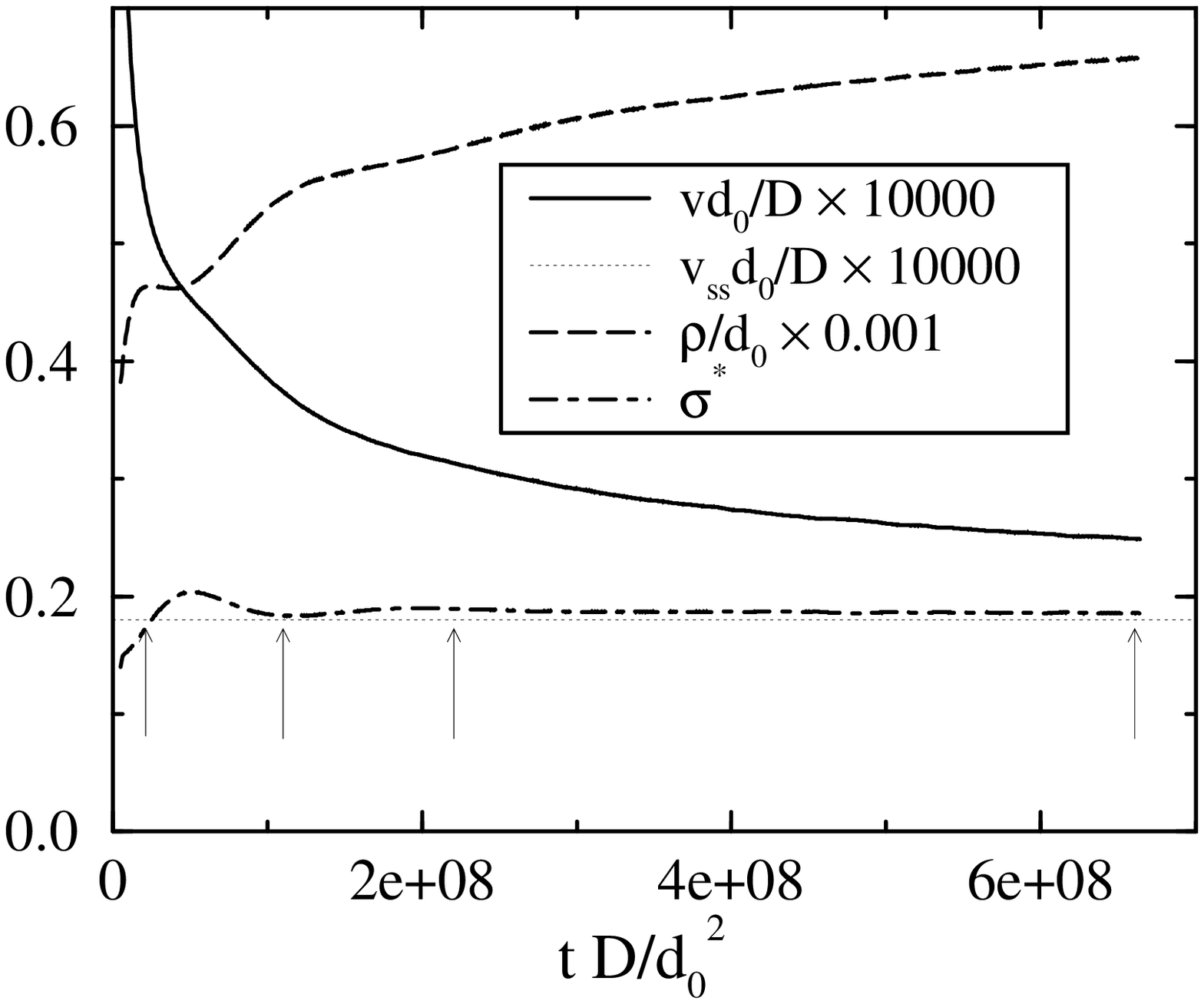,width=.32\textwidth}}
\smallskip
\caption{
Result of a 3-d dendrite growth simulation
for $\Delta=0.05$ and $\epsilon_4 = 0.025$ with 
snapshots of the 3-d structure (top) at the times corresponding
to the arrows. This run took 6 hours 
on 64 processors of the CRAY T3E at NERSC and used
up to $5\times 10^6$ walkers.}
\label{figden}
\end{figure}

To illustrate our method, we focus here 
on the initial stage of dendritic solidification,
during which four (six) primary arms in 2-d (3-d)
emerge from a structureless nucleus, as shown
in the example 3-d run of Fig. \ref{figden}. 
While this transient regime
has recently been investigated numerically in 2-d and
experimentally in 3-d \cite{Proetal99}, 
it has not yet been explored by simulations in 3-d. 

The present simulations started from a small spherical solid nucleus
with a uniformly undercooled temperature $u=-\Delta$,
and fully exploited a cubic symmetry 
(i.e. $a(\hat n)\equiv (1-3\epsilon_4)
[1-4\epsilon_4/(1-3\epsilon_4)(n_x^4+n_y^4+n_z^4)]$
with the Cartesian axes
defined parallel to the [100] directions) to
reduce simulation time. 
The value $\epsilon_4=0.025$ that
corresponds to the experimentally estimated anisotropy value for
pivalic acid (PVA) \cite{Musetal} was used in all the simulations 
reported here. Results for other anisotropies
and that pertain to the 3-d morphology of the dendrite tip will be
discussed elsewhere. To obtain quantitative data
on the growth transients,
we recorded the arm length $L(t)$, the tip velocity
$v(t)=\dot L(t)$, the tip radius of curvature $\rho(t)$, 
and the total volume of solid $V_s(t)$.
In Fig. \ref{figden}, we show $v(t)$ and $\rho(t)$
for a 3-d run at $\Delta=0.05$ together with the
time-dependent tip selection parameter 
$\sigma^*(t)=2d_0D/[\rho^2(t) v(t)]$ and the steady-state
velocity $v_{ss}$ calculated by a boundary integral
method. Two results are particularly noteworthy. 
Firstly, $\sigma^*(t)$ becomes
essentially constant as soon as the arms have emerged from 
the spherical seed, whereas both $v(t)$ and 
$\rho(t)$ are far from their steady-state values.
Physically, this is a direct consequence of the fact that
$v(t)$ and $\rho(t)$ evolve slowly on the tip diffusion
time scale $\rho^2/D$ where $\sigma^*$ is established.
Secondly, we find that, as the undercooling is lowered,
the volume of the dendrite (or its area in 2-d) 
approaches that of a sphere (circle) growing at the
same undercooling. The latter is readily obtained
from Zener's well-known similarity solution, which
yields that the radius of a $d$-dimensional sphere
grows as $\sqrt{4p_dDt}$, where the Peclet number
$p_d(\Delta)$ is implicitly defined by
$\Delta = p_d^{d/2} \exp(p_d) \int_{p_d}^\infty s^{-d/2} e^{-s} ds$.
Both in 2-d and 3-d, the volume of the dendrite grows
slightly faster than the one of the sphere, but for
the lowest undercoolings we could attain, the final
volume differed by only $20\%$ from this prediction,
even though the arms were already very well developed.

\begin{figure}
\centerline{
\psfig{file=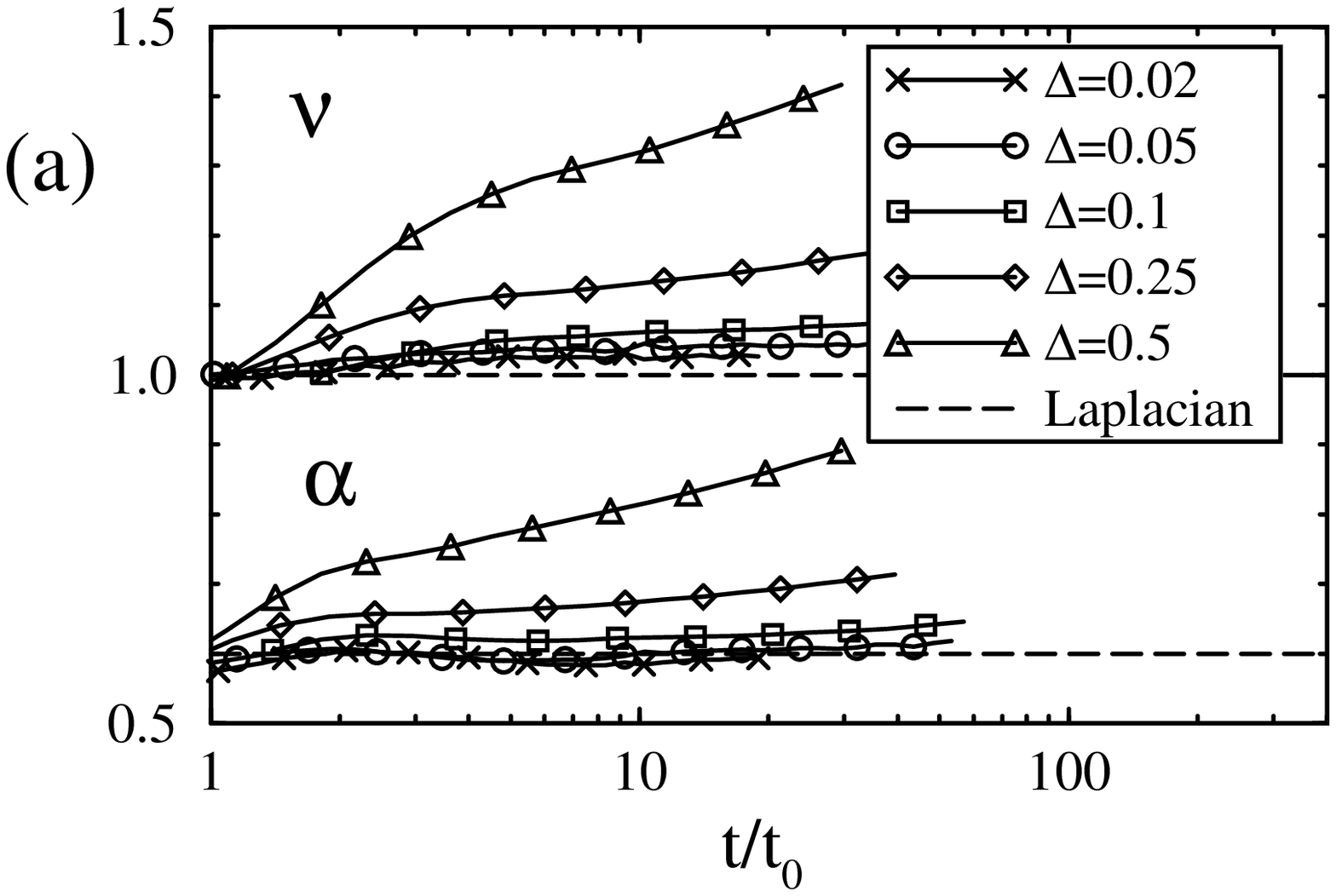,width=.38\textwidth}}
\centerline{
\psfig{file=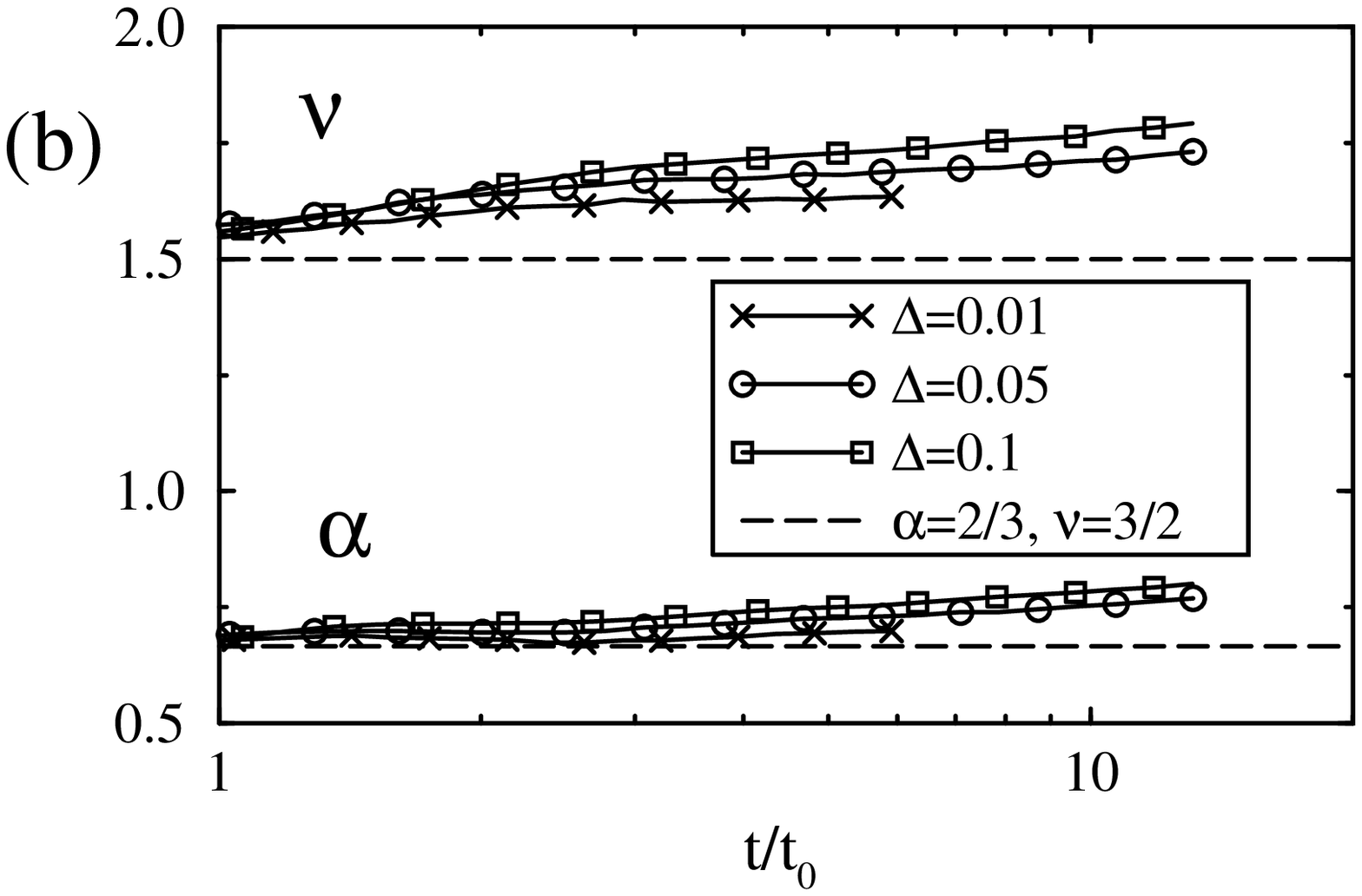,width=.38\textwidth}}
\caption{Functions $\alpha(t)$ and $\nu(t)$ vs scaled
time $t/t_0(\Delta)$ in 2-d (a)
and 3-d (b). In order to show all curves on the same plot,
$t_0(\Delta)$ was defined to be the time
at which the tips are ahead of the grooves by about two
tip radii. In order of decreasing $\Delta$, 
$t_0D/ d_0^2=2.29 \times 10^4$, $9.03 \times 10^5$,
$4.36 \times 10^7$, $5.35 \times 10^8$, and
$1.39 \times 10^{10}$ in 2-d, and
$4.18 \times 10^6$, $5.58 \times 10^7$ and,
$1.17 \times 10^{10}$ in 3-d.}
\label{figexpo}
\end{figure}

The above observations are in good agreement with 
theoretical expectations for 2-d growth
transients. In particular,
Almgren {\it et al.} have analyzed the related problem 
of anisotropic Hele-Shaw flow (i.e. Laplacian growth) at 
constant flux \cite{Almetal}. Using an exact solution for
the Laplacian field around a 
cross and exploiting the 
constancy of $\sigma^*$, they constructed a 
self-affine scaling shape for the arms. The length and width
of this shape grows as $t^\alpha$ and $t^\beta$, respectively,
with $\alpha=3/5$ and $\beta=2/5$.
As subsequently remarked by Brener \cite{Bre}, 
the diffusion equation can be replaced by Laplace's equation on
the scale of the dendrite as long as 
$L/\sqrt{Dt}\ll 1$, and hence 2-d growth 
transients should obey this scaling
with a flux set by the diffusive far field of
Zener's similarity solution in the low
undercooling limit.
In Fig. \ref{figexpo}a, we plot the 
functions $\alpha(t)=d(\ln L)/d(\ln t)$ and
$\nu(t)=d(\ln V_s)/d(\ln t)$. For an exact
Laplacian scaling in 2-d, $\alpha(t)=3/5$ and
$\nu(t)=1$. With decreasing undercooling both 
curves become flatter and indeed approach the expected Laplacian 
scaling. The slow rise with time of both curves
can be attributed to diffusive corrections
to the Laplacian scaling due to 
the slow increase in time of $L/\sqrt{Dt}$.
Recently, Provatas {\it et al.} have reported 
scaling exponents that differ from the Laplacian
prediction and that appear to be independent
of $\Delta$ for small $\Delta$ \cite{Proetal99}. 
We note, however, that they used the distance from
the tip to the time-dependent base
(where the dendrite shaft is narrowest) to scale their
results instead of $L(t)$. Since 
$L(t)$ is the only relevant scaling length for both the shape and
the diffusion field in the Laplacian limit, we believe that these
exponents are spurious. When $L(t)$ is used as a scaling
length, our results are consistent with an approach
to Laplacian scaling in the limit of vanishing undercooling.

It is simple to heuristically generalize some
of the above ideas to 3-d. If we assume, as a
reasonable first approximation, that the arm  
shape is axisymmetric and has a scaling form
$r(x,t)=t^\beta\tilde r\left(x/t^\alpha\right)$, where
$x$ is the growth direction,
the constancy of $\sigma^*$ imposes that $4\beta-\alpha-1=0$.
Furthermore, assuming that the
volume of the dendrite grows approximately
as the one of the 3-d similarity solution, 
we have in addition $\nu=\alpha+2\beta=3/2$. These two
conditions yield $\alpha=2/3$ and $\beta=5/12$.
Fig. \ref{figexpo}b shows the functions $\alpha(t)$
and $\nu(t)$, defined as before, in 3-d. As in 2-d,
the curves approach the predicted exponents with
decreasing undercooling, but the differences remain
larger than in 2-d even for the lowest undercooling.
This can be partly accounted for by the
fact that the tip velocity, and hence also
$L/\sqrt{Dt}$, is much larger in 3-d than in 2-d at
equal undercoolings. We must emphasize
that in the absence of an exact 3-d Laplacian solution,
and in view of the above assumptions,
no claim is made here that these 3-d exponents are exact or that
a scaling regime exists asymptotically at small undercooling
in 3-d. We content ourselves with the fact that they
describe reasonably well our present simulations. 

In conclusion, we have presented a novel computational
approach that can resolve accurately the details
of a complex branched structure and its large scale
surrounding diffusion field. The method can be combined with
many of the existing front tracking methods
and should be applicable to a wide
range of diffusion-limited pattern forming systems.
Furthermore, we have demonstrated its feasibility in
the non-trivial test case of dendritic growth in
a range of parameters previously unreachable in 3-d.

This research is supported by U.S. DOE Grant 
No. DE-FG02-92ER45471 and benefited from 
computer time allocation at NERSC and NU-ASCC.
We thank Flavio Fenton for help with 3-d visualization
of our simulations and Vincent Hakim 
for fruitful conversations.

\end{document}